\documentclass[sigconf]{acmart}
\pdfoutput=1
\usepackage{cite}
\usepackage{amsmath,amssymb,amsfonts}
\usepackage{graphicx}
\usepackage{textcomp}
\usepackage{xcolor}
\usepackage{todonotes}
\usepackage{pifont}
\usepackage{amsmath}
\usepackage{appendix}
\usepackage{enumitem}
\usepackage{natbib}
\usepackage{bm}
\usepackage{afterpage}
\usepackage[noend]{algpseudocode}
\usepackage{algorithm}
\usepackage{subfig}
\usepackage{float}
\usepackage{caption}

\def\acmBooktitle#1{\gdef\@acmBooktitle{#1}}
\acmBooktitle{Proceedings of \acmConference@name
       \ifx\acmConference@name\acmConference@shortname\else
         \ (\acmConference@shortname)\fi}

\begin{document}
\title{HAP-SAP: Semantic Annotation in LBSNs using Latent Spatio-Temporal Hawkes Process}
\author{Manisha Dubey}
\email{cs17resch11003@iith.ac.in}
\affiliation{%
  \institution{Computer Science and Engineering \\
IIT Hyderabad}
}

\author{P.K. Srijith}
\email{srijith@cse.iith.ac.in}
\affiliation{%
  \institution{Computer Science and Engineering \\ IIT Hyderabad}
}

\author{Maunendra Sankar Desarkar}
\email{maunendra@cse.iith.ac.in}
\affiliation{%
  \institution{Computer Science and Engineering \\ IIT Hyderabad}
}

\renewcommand{\shortauthors}{Manisha Dubey, P.K. Srijith, Maunendra Sankar Desarkar}

\begin{abstract}
 The prevalence of location-based social networks (LBSNs) has eased the understanding of human mobility patterns. Knowledge of human dynamics can aid in various ways like urban planning, managing traffic congestion, personalized recommendation etc. These dynamics are  influenced by factors like social impact, periodicity in mobility, spatial proximity, influence among users and semantic categories etc., which makes location modelling a critical task. However, categories which act as semantic characterization of the location, might be missing for some check-ins and can adversely affect modelling the mobility dynamics of users. At the same time, mobility patterns provide a cue on the missing semantic category. In this paper, we simultaneously address the problem of semantic annotation of locations and location adoption dynamics of users. We propose our model HAP-SAP, a latent spatio-temporal multivariate Hawkes process, which considers latent semantic category influences, and temporal and spatial mobility patterns of users.  The model parameters and latent semantic categories are inferred using expectation-maximization algorithm, which uses Gibbs sampling to obtain posterior distribution over latent semantic categories. The inferred semantic categories can  supplement our model on predicting the next check-in events by users. Our experiments on real datasets demonstrate the effectiveness of the proposed model for the semantic annotation and location adoption modelling tasks. 
\end{abstract}


\copyrightyear{2020}
\acmYear{2020}
\setcopyright{rightsretained}
\acmConference[SIGSPATIAL '20]{28th International Conference on Advances in Geographic Information Systems}{November 3-6, 2020}{USA}
\acmBooktitle{28th International Conference on Advances in Geographic Information Systems (SIGSPATIAL '20), November 3-6, 2020, Seattle, WA, USA}
\acmDOI{10.1145/3397536.3422233}
\acmISBN{978-1-4503-8019-5/20/11}
\maketitle
\section{Introduction}
Online social networks (OSN) provide a platform to share information across the network of users. The advent of location acquisition technologies like GPS has motivated the development of location-based social networks (LBSNs). LBSNs provide an online social networking platform where users can mark their visit (`check-in’) to location of interest known as  \textit{point-of-interest} (POI) such as any hotel, restaurant, theater, etc., share photos, and mark their location. Each check-in specifies the time of visit along with location of the place visited by user as POI. Some platforms also capture user preference and interests in terms of category associated with the location visited. LBSNs such as Foursquare, Gowalla, BrightKite etc. are a rich source of information about various locations of interest, social network connections among people, user check-ins and their mobility patterns.   Widespread adoption and use of location-based social networks is evident from the fact that social networking giants like Facebook and Twitter have incorporated check-in facility into their platforms.
Modelling LBSNs can help in modelling human dynamics in the real world which can be helpful in making better personalized recommendations and context-aware systems. Also, it can be useful in a myriad of applications like urban planning, transport management etc. 
Location-based social networks being an impression of human movement and dynamics is affected by various factors like social influence, geographical and temporal patterns, periodicity in human mobility.  It was found that human mobility patterns experience a combination of periodic movements that are geographically limited and seemingly random jumps correlated with their social networks \citep{paper2}. 
In addition to the above mentioned challenges,  it was found that 30\% of the locations in Foursquare and Whrrl lack meaningful textual descriptions representing their semantic categories \citep{paper3}. Discovering these categories can be quite useful in modelling LBSNs in a better way. 

\begin{figure}
    \begin{center}
    \includegraphics[width=0.90\columnwidth, height=0.25\textwidth]{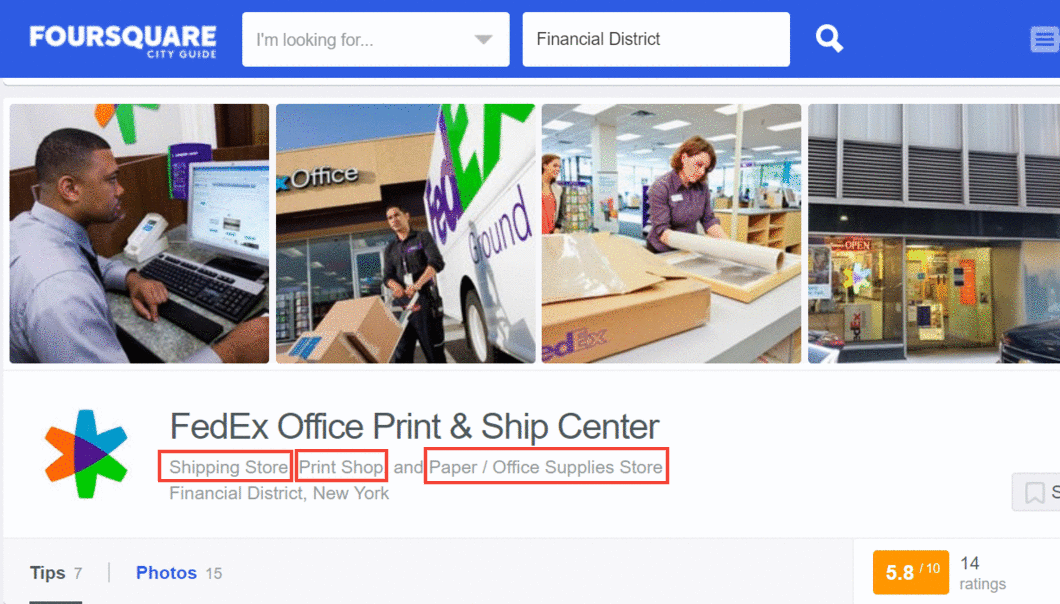}
    \caption{A snapshot of description of a venue in Foursquare. The venue is associated with multiple categories (marked with red). This emphasizes that a venue based check-in may be insufficient to capture user intent. [Source: \href{https://foursquare.com/v/fedex-office-ship-center/4dd2e7f9e4cd1b19130c93e2}{Foursquare}]}
    \label{fig:foursquare_event}
    \end{center}
\end{figure}

Traditional methods have addressed this issue from the perspective of missing categories for venues. However, we propose that a venue may be associated with multiple categories but intent of user is captured by the category being visited at a particular time. For instance, as shown in Figure \ref{fig:foursquare_event}, a venue is associated with multiple categories like \textit{Shipping Store}, \textit{Print Shop} and \textit{Office Supply Store}. A user may be interested in only one of these categories at a time. 
Considering this, we propose that it is imperative to associate a category to an event while doing semantic annotation in order to capture the real intent of user in an efficient way. This motivates development of a comprehensive model which can simultaneously discover  semantic categories of each event and model mobility dynamics of users.

In this paper, we propose a  model based on Hawkes process for semantic annotation of places (HAP-SAP).  In particular, we propose a latent multivariate spatio-temporal Hawkes process to model semantic annotation and location adoption dynamics in location-based social networks. This allows to build a comprehensive model to incorporate temporal factors, geographical influence and user-interests. Each check-in category is considered as a mark associated with the event. We model the missing category associated with the event as latent mark. We employ expectation-maximization procedure to infer the missing categories and learn model parameters.  Additionally, the learnt parameters and inferred categories are used to understand mobility dynamics and predict future check-ins. Our key contributions in this paper are:
\begin{itemize}
    \item Modelling LBSNs using multivariate spatio-temporal Hawkes process with missing categories as latent marks  
    \item Joint discovery of latent semantic categories and location adoption behaviour modelling
    \item Expectation-Maximization with Gibbs sampling under the framework of Hawkes process to estimate  missing categories in LBSNs and model parameters.
    \item Experiments showing the effectiveness of our model for semantic annotation and location adoption modelling
\end{itemize}

\section{Related Work}
A fundamental problem associated with LBSNs is that they suffer from the problem of missing data where places lack any meaningful textual description. The problem of finding missing labels for the POI is known as semantic annotation of location in the domain of LBSNs. Previous work poses the problem of semantic annotation as multi-label classification problem \citep{paper3} where a binary SVM classifier is learnt for each tag. They considered features like population  features  (e.g.,  number  of  unique visitors)  and  temporal  features  (e.g.,  distribution  of  check-in time)  as  semantic  descriptions  of  specific  places. To capture implicit relatedness, they built a network of related places and found relatedness using random walk and restart technique.
In \citep{paper17}, authors have used a semi-supervised learning framework based on graph embedding for semantic annotation. They have learnt user embedding from a user-tag bipartite graph. Place embedding is represented as the centroid of the vectors of its check-in users. The authors  in \citep{paper16}  used a probabilistic topic model considering the factors like user interests, temporal and spatial pattern, rating score to find category-aware and sentimental tags. A recent work in \citep{paper32} extracts user similarities and performs multi-label semantic annotation with extreme learning machine. 

Point processes have been found quite useful to model location adoption dynamics in LBSNs. A doubly stochastic periodic point process was proposed in \citep{paper21} to predict time and location of check-ins. The effectiveness of Hawkes  process (HP) for modelling location adoption dynamics is demonstrated in \citep{paper9}. They make use of historical check-in events and influence between users to effectively model location adoption dynamics.  Temporal point processes such as Hawkes process were adapted to model for missing marks or events in an event history data~\citep{paper34,paper35,paper38,paper7,paper39}.    They largely consider modelling the data with intermittent temporal events and do not consider complex spatio-temporal patterns. A related work  is \citep{paper36} where authors use a spatio-temporal Hawkes process to model and infer interacting pairs of users. However, these models are not applicable to the semantic annotation task in LBSNs, where categories or marks are   partially observable. Hence, we propose a latent spatio-temporal Hawkes process  with an effective inference mechanism to address the  problem of semantic annotation in LBSNs. The capability of the HP to consider historical check-in events and  influences between marks can help to correctly infer missing categories associated with events.
\vspace{-1mm}
\begin{figure*}[ht]
    \begin{center}
    \includegraphics[width=0.7\paperwidth, height =0.25\paperwidth]{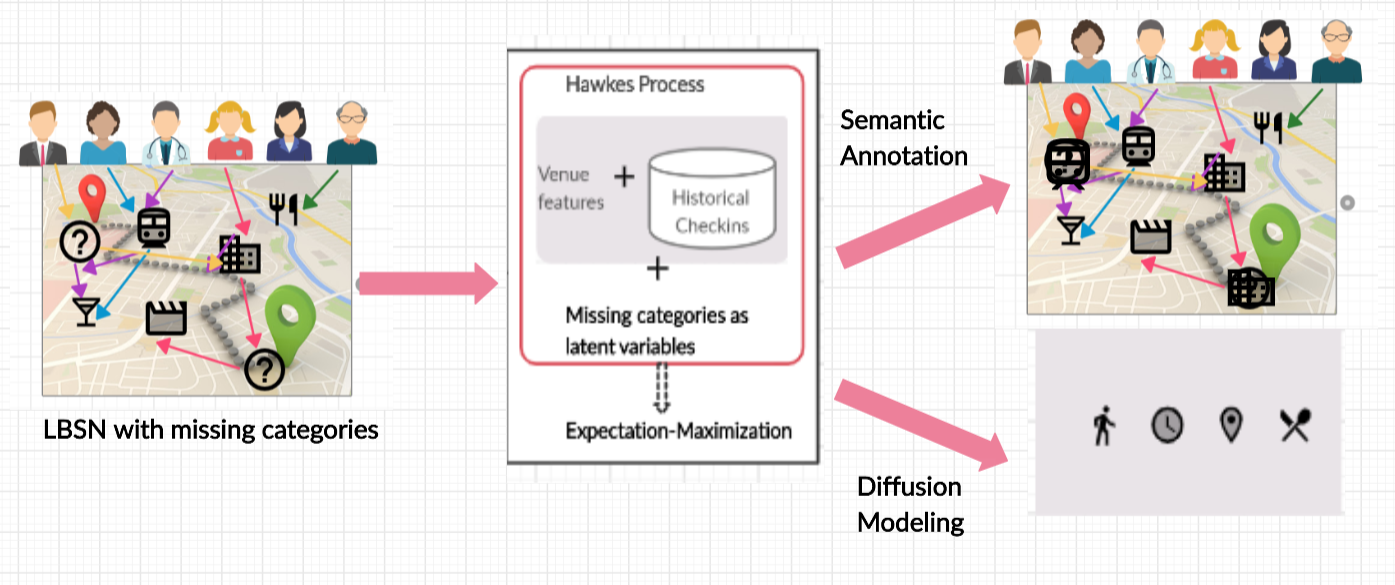}
    \caption{Framework for HAP-SAP model: Input for the model is a location based social network dataset with missing categories (marked with '?'). We treat missing categories as latent variables and use Hawkes process to model check-ins. Output contains a complete dataset without any missing categories. Also, we predict future check-ins using learnt parameters and complete dataset }
    \label{fig:framework}
    \end{center}
\end{figure*}
\section{Problem Definition}
\label{problem_defn}
Consider a location-based social network $G$.  An activity of marking a user's visit to a location is known as \textit{Check-in}. A check-in is also known as \textit{Event}. A location with an associated value of latitude and longitude which may be of interest of users is known as \textit{Point-of-Interest} (POI), which we also refer as \textit{venue}. Such a place is also characterized with a category which represents semantic meaning of the type of place. 
Each check-in is associated with user-id $u$, venue-id $v$, timestamp $t$, location $l$ and category $c$. Each location $l$ is a point of interest with a unique identifier and hence has a pair of latitude $(x)$ and longitude $(y)$ associated with it. 

We will define a LBSN $G$ with $M$ users represented as $U= \{u_1, \ldots, u_{M}\}$, $W$ locations as $L = \{l_1, \ldots, l_W\}$, P POIs (venues) as $V = \{v_1, \ldots, v_{P}\}$ and $K$ categories as $C = \{c_1, \ldots, c_{K}\}$. Each POI is associated with any of these categories. 
Assuming each check-in (event) to be $e_n$ and the number of check-ins be $N$. $G$ contains $\mathbf{E} = \{u_n, v_n, t_n, l_n, c_n\}_{n=1}^{N}$ check-ins where each $l_n$ consists of a pair $\{x_n, y_n\}$ representing latitude and longitude corresponding to location $l_n$, $u_n$ representing user-id, $v_n$ represents the venue-id, $t_n$ representing timestamp  and $c_n$ representing category associated with the $n^{th}$ check-in. $H_u$ represents the check-in event history for user $u$ which refers to  the entire set of events which have happened before a particular event for the user $u$. 

In the problem addressed in this paper,  categories are missing for some of the check-in events. Let us assume that out of $N$ number of check-in events, $N_c$ number of events have  observed categories. We can define set of events with observed categories  $\mathbf{E}_c$ as $\{u_n, v_n, t_n, l_n, c_n\}_{n=1}^{N_c}$.  Also, let $N_z$ be the number of events with unobserved categories and set of events with unobserved categories be $\mathbf{E}_z = \{u_n, v_n, t_n, l_n \}_{n=1}^{N_z}$.  We define $\mathbf{E} =\mathbf{E}_c \cup \mathbf{E}_z$  with N events where  $N = N_c + N_z$. Thus,
\setlength{\belowdisplayskip}{1pt} \setlength{\belowdisplayshortskip}{1pt}
\setlength{\abovedisplayskip}{1pt} \setlength{\abovedisplayshortskip}{1pt}
\begin{equation*}
    \mathbf{E} = \{u_n, v_n, t_n, l_n, c_n\}_{n=1}^{N_c} \cup \{u_n, v_n, t_n, l_n \}_{n=1}^{N_z}
\end{equation*}
We will denote the set of  unobserved categories as $\mathbf{z} = \{z_n\}_{n=1}^{N_z}$. Our goal is to detect the unobserved categories  $\mathbf{z}$  associated with the check-in events. We  also aim to predict future occurrences of check-in events along with inferring the missing categories. 

\section{BACKGROUND}
In this section, we will provide a preliminary discussion on Hawkes process and their use in modelling spatio-temporal data. 

\subsection{Hawkes Process}
Point processes are useful to model the distribution of points over some space and are defined using an underlying intensity function.  
A Hawkes process \citep{paper24} is a point process with self-triggering property i.e occurrence of previous events trigger occurrences of future events. 
Conditional intensity function for univariate Hawkes process is defined as
 \[ \lambda(t) = \mu + \sum_{t_k < t}k(t - t_k) \]
 where 
 $\mu$ is the base intensity function and $k(\cdot)$ is the triggering kernel function capturing the influence from previous events. The summation over $t_k < t$ represents all the effect of all events prior to time $t$ which will contribute in computing the intensity at time $t$. 
 Hawkes process has been used in earthquake modelling \citep{paper20}, crime forecasting \citep{paper19} and epidemic forecasting \citep{paper18}. \\

\subsection{Multivariate Hawkes Process}
Events are most often associated with features other than time such as categories or users in LBSNs. Such features are known as marks.
The multi-variate Hawkes process \citep{paper25} is a multi-dimensional point process that can model time-stamped events with marks. It allows explicit representation of marks through the $i^{th}$ dimension of the intensity function and can capture influences across these marks. The intensity function associated with the $i^{th}$ mark is 
\begin{displaymath}
	\lambda_i(t) = \mu_i + \sum_{t_k < t} \alpha_{ii_k} k(t - t_k)
\end{displaymath}
where $\mu_i> 0$ is the base intensity of $i^{th}$ mark. We consider that previous event $k$ is associated with a mark $(i_k)$ and is treated as a dimension in Hawkes process. The intensity at time $t$ for a mark $i$ is assumed to be influenced by all the events happening before $t$ at time $t_k$ and mark $i_k$. The influence of mark $i_k$ on some mark $i$ is given by $\alpha_{ii_k}$. This models the mutual excitation property between events with different marks. With respect to our application, the mark would represent the category associated with the check-in event.

\subsection{Spatio-Temporal Hawkes Process}
A spatio-temporal point process models the occurrence of points over a spatio-temporal domain  \citep{paper26}.  They can be quite useful to model LBSNs, where check-in events are associated with both a location (spatial coordinate) and time. Here, the conditional intensity function for an event with mark $i$ at location $l$ and $t$ can be defined as

\begin{displaymath}
	 \lambda_i(l,t) = \mu_i + \sum_{t_k < t} \alpha_{ii_k} k(l - l_k, t - t_k)
\end{displaymath}
Here, triggering kernel not only captures influence in terms of time, but also in terms of location.

\section{Proposed Model}
In this section, we discuss the proposed model HAP-SAP which is a spatio-temporal Hawkes process approach to detect the missing categories. Figure ~\ref{fig:framework} displays the outline of our proposed approach. 
As discussed above, human mobility patterns are affected by various factors like periodicity in visiting some locations, burstiness for certain occasions and influence of one category over another. Hence, we need to build a model  which considers temporal and spatial patterns and is affected by historical events as well. Due to these reasons, we use a spatio-temporal Hawkes process to model the location based social networks where categories are considered as marks associated with the event.   

In the following discussion we assume categories associated with all the events are observed ($z_n$ are observed variables). Hence, $\mathbf{E}$ dataset consists of events with categories which are observed. We extend the model to the case with unobserved categories in  
Section \ref{stlhp}. Given $\mathbf{E}$, we can express the intensity function for a user $u_n$ visiting location $l_n$ with the category $c_n$ at time $t_n$
considering the multivariate spatio-temporal Hawkes process model for LBSNs as following - 
\begin{equation}
     \lambda_{c_n}^{u_n}(t_n,l_n) = \mu_{c_n} + \sum_{t_k < t_n} \mathbb{I} (u_k = u_n) \alpha_{c_{n} c_{k}} k(t_n-t_k, l_n-l_k) 
\end{equation}

Here, $\alpha$ matrix represents the latent influence between the categories. The matrix element $\alpha_{c_{n} c_{k}}$ gives the influence of $c_k$ in producing a category $c_n$ (the likelihood that user will visit category $c_n$ after visiting $c_k$). This combined with the time and location at which the previous event happened determines the influence of the previous event on the current event. 
The triggering kernel is assumed to be a product of separate kernels over space and time. 
\begin{equation}
\begin{split}
& k(t_n-t_k, l_n-l_k) = k(t_n-t_k) \times k(l_n-l_k) \\
& \text{where} \\
& k(t_n-t_k) = exp(-\eta(t_n-t_k)) \\
& k(l_n-l_k) = exp \Big(- \dfrac{|| l_n-l_k ||}{2h} \Big)
\end{split}
\end{equation}
where $\eta$ and $h$ are temporal and kernel parameters, known as decay and bandwidth respectively.
\\
The base intensity influences the arrival of events due to exogenous factors. In a standard Hawkes process model, base intensity is constant and learnt from the data. In our model, we propose base intensity to be a function of venue features.

Motivated from the work proposed in \citep{paper3}, we consider the following features:
\begin{itemize}[partopsep=6pt,topsep=6pt,parsep=6pt]
    \item \textbf{Check-in day:}
    Each venue has its own categorical relevance over different days of a week. For example, a venue with category as 'professional' may have more check-ins on weekdays than weekends. On the contrary, there may be some venues with 'restaurant' category which may have more number of check-ins on weekends. Hence, we represent each venue with a feature vector where each entry represents number of check-ins on that venue in a particular day of the week. In this way, we use venue feature to find base intensity for a category.  
    \item \textbf{Check-in time:}
    A venue may be more checked-in in morning than at night. To capture this, we represent each venue on a 24-hour scale. Each venue is represented as a function of number of check-ins on hour scale. This feature contributes to the base intensity for a category based on the check-in time to the venue. To reduce the number of parameters of check-in time, we have performed binning of 24-hour scale into four bins.
    \end{itemize}
We propose the base intensity associated with a  category $c$ to be the following:
\begin{equation}
\begin{split}
    \mu_{c} &= \mu^{day}_{c} + \mu^{hour}_{c}
    \\
    \mu^{day}_{c} &= \exp{(\mathbf{w}_{c}^{day}\cdot \mathbf{x}^{day})} \\
    \mu^{hour}_{c} &= \exp{(\mathbf{w}^{hour}_c\cdot \mathbf{x}^{hour})} 
    \end{split}
    \end{equation}
\begin{itemize}
    \item $\mathbf{x}^{day}$ represents feature of venue as distribution of check-ins in a week
    \item $\mathbf{x}^{hour}$ represents feature of venue as distribution of check-ins in hour-scale 
\end{itemize}
Using the combination of these features, we learn different base intensities for different categories based on venue. 
We will learn $\mathbf{w}^{day}_c$ and $\mathbf{w}^{hour}_c$ during parameter estimation step.  

The parameters $(\boldsymbol{\theta})$ of the intensity functions listed in Table~\ref{tab:param} are estimated by maximizing the likelihood of observing the check-in events, $p(\mathbf{E}|\boldsymbol{\theta})$. This likelihood is given as
\begin{equation}
L(\boldsymbol{\theta}) = \prod_{n=1}^{N} \lambda_{c_n}^{u_n}(t_n,x_n, y_n) \exp(-\sum_{u=1}^{M} \sum_{c=1}^{K}\int_{T} \int_{X}\int_{Y} \lambda_{c}^u(t,x,y)dydxdt
\end{equation}
where the first product term represents the instantaneous probability of occurrence of events while the exponential term represents the probability that no event happens outside the check-in events. 

The parameters are estimated by maximizing the log likelihood which can be written as:
\begin{equation}
\begin{split}
LL(\boldsymbol{\theta}) &=  \sum_{n=1}^{N} \log \lambda_{c_n}^{u_n}(t_n,x_n, y_n) - \\
& \sum_{u=1}^M \sum_{c=1}^{K}  \int_{T_{min}}^{T_{max}} \int_{X_{min}}^{X_{max}}\int_{Y_{min}}^{Y_{max}} \lambda_{c}^u (t,x,y)dydxdt 
\end{split}
\end{equation}

\begin{equation} 
\begin{split}
 & =  \sum_{n=1}^{N} \log \lambda_{c_{n}}^{u_n}(t_n, x_n, y_n) - \big ( \sum_{c=1}^{K} \mu_{c} \big) MTXY - \\
& \sum_{\underset{u=1}{c=1, n=1}}^{M,K,N}\! \sum_{k=1}^{n-1} \int_{t_{n-1}}^{t_n}\!\!\!\int_{x_{n-1}}^{x_n}\!\!\! \int_{y_{n-1}}^{y_n} \!\!\!\!\!\!\!\! \mathbb{I} (u_k = u) \alpha_{c c_{k}} k(t-t_k,\! x-x_k,\! y - y_k)dydxdt
\end{split}
\end{equation}
Here $T = T_{max} - T_{min}$, $X = X_{max} - X_{min}$ and $Y = Y_{max} - Y_{min}$ are the ranges of time and location coordinates.

\subsection{Spatio-Temporal Latent Hawkes Process}
\label{stlhp}

In many real world applications, categories might be missing for some check-ins. This brings in additional complexity in modelling the location adoption dynamics of users. We consider missing categories as latent variables in the spatio-temporal Hawkes process model. This allows us to use concepts from latent variable modelling in detecting the missing categories \citep{paper31}. In particular, we use an  expectation-maximization algorithm to  detect the missing categories and learn the parameters of the model. In the further sub-sections, we propose the components of our model including characterization of intensity function along with parameter estimation and inference and thereby build a complete model. Further, we will explain generative process for our model.

We modify the intensity function to account for the missing categories or marks. The categories are observed for some check-in events  while for others they are latent. The missing categories affects the intensity function calculation for future check-in events. The latent category for a check-in also depends on the past categories and the influence of past check-in events through the intensity function. This interplay between the categories will be helpful in detecting the missing categories. The intensity function has to be modified to consider both the observed and the latent variables.  

As defined in section \ref{problem_defn}, we assume $\mathbf{E}$ to consist of all the check-in details except the details of the missing categories. Let $N$ be the total number of check-in events, with $N_c$ number of events with   observed categories and  $N_z$ number of events with unobserved categories. We denote the unobserved categories for an event as $z_n$. For modelling the missing categories in our problem, we split the intensity function for an event into two cases - \\
\textbf{\textit{Case 1:}} The intensity function for the first case where the category is observed for the $n^{th}$ event is defined as   
\begin{equation}
\begin{split}
\lambda_{c_n}^{u_n}(t_n,l_n) & = \mu_{c_n} + \!\!\!\!\! \sum_{t_k < t :  c_k \text{ is observed}} \!\!\!\!\! \mathbb{I} (u_k = u_n) \alpha_{c_n c_k} k(t_n-t_k, l_n-l_k) + \\
&\sum_{t_k < t : c_k\text{ is unobserved}} \mathbb{I} (u_k = u_n) \alpha_{c_n z_k} k(t_n-t_k, l_n-l_k) 
\end{split}
\end{equation}
The second term considers the scenario for which previous check-in events have observed categories. While the third term considers the case that previous check-in events have unobserved categories. 
\\ \\
\textbf{\textit{Case 2: }} The following intensity function considers the second case where the category associated with the $n^{th}$ event is unobserved. 
\begin{equation}
\begin{split}
\lambda_{z_n}^{u_n}(t_n,l_n) & = \mu_{z_n} + \sum_{t_k < t :  c_k \text{ is observed}} \alpha_{z_n c_k} k(t_n-t_k, l_n-l_k) + \\
&\sum_{t_k < t : c_k\text{ is unobserved}} \alpha_{z_n z_k} k
(t_n-t_k, l_n-l_k) 
 \end{split}
\end{equation}

Considering the newly defined aforementioned intensity function with some values assigned to the latent categories (conditioned on $\mathbf{z}$),  the likelihood $p(\mathbf{E}|\mathbf{z},\boldsymbol{\theta})$ is defined as 
\begin{eqnarray}
\label{eqn:cond_like}
p(\mathbf{E}|\mathbf{z},\boldsymbol{\theta}) = \prod_{n=1}^{N_c} \lambda_{c_n}^{u_n}(t_n,x_n, y_n) \prod_{n=1}^{N_z} \lambda_{z_n}^{u_n}(t_n,x_n, y_n) \nonumber \\  \exp(-\sum_{u=1}^{M} \sum_{c=1}^{K}\int_{T} \int_{X}\int_{Y} \lambda_{c}^u(t,x,y) dydxdt 
\end{eqnarray}

\subsection{Parameter Estimation and Inference}

The likelihood function consists of observed data $\mathbf{E}$, missing categories $\mathbf{z}$ and parameters to be estimated as $\boldsymbol{\theta}$. A maximum likelihood approach can be used to learn the parameters if all the categories were observed. However, as some categories are latent ($\mathbf{z}$), we resolve to an iterative procedure called expectation-maximization algorithm to  estimate the latent variables and parameters.

\subsubsection{\textbf{Expectation Step}}
\label{param_est_estep}
The expectation step finds the expected value of the joint likelihood over observed and latent variables with respect to the posterior distribution over the latent variables. 
The latent variables which are missing categories are discrete in nature and can take $K$ possible values. We assume the latent variables are independently and identically distributed as multinoulli or categorical distribution (prior distribution) with $K$ parameters.
\begin{equation}
p(\mathbf{z})  = \prod_{i=1}^{N_z} p(z_i) = \prod_{i=1}^{N_z} Cat(z_i;p)
\end{equation}

Now, the posterior distribution is obtained by combining the prior with the likelihood defined in Equation \eqref{eqn:cond_like}.
\begin{equation}
 p (\mathbf{z} | \mathbf{E}, \boldsymbol{\theta}) = \frac{ p(\mathbf{E} | \mathbf{z}, \boldsymbol{\theta}) p(\mathbf{z})}{p(\mathbf{E})} 
\end{equation}
However, due to non-conjugacy between likelihood and prior, the posterior can not be obtained in a closed form. Hence, the expectation of the joint likelihood with respect to posterior can not be computed analytically.  
\begin{equation}
\Tilde{L}(\boldsymbol{\theta}) = \mathbb{E}_{p(\mathbf{z}|\mathbf{E},\boldsymbol{\theta})} [\log p(\mathbf{E},\mathbf{z} | \boldsymbol{\theta})] = \mathbb{E}_{p(\mathbf{z}|\mathbf{E},\boldsymbol{\theta})} [ \log \big (p(\mathbf{E} | \mathbf{z}, \boldsymbol{\theta}) p(\mathbf{z})\big )]
\end{equation}
For finding expected log joint likelihood, we will use Monte Carlo approximation. We get the samples from the posterior using Gibbs sampling ($\mathbf{z}^{s}$) and use these samples to approximate the expected log joint likelihood.
\begin{equation}
\begin{split}
\mathbb{E}_{p(\mathbf{z}|\mathbf{E},\boldsymbol{\theta})} [ \log p(\mathbf{E} | \mathbf{z}, \boldsymbol{\theta}) & + \log p(\mathbf{z})]  = \frac{1}{N} \sum_{s=1}^S  [ \log p(\mathbf{E} | \mathbf{z}^{s}, \boldsymbol{\theta}) + \\
& \log p(\mathbf{z}^{s})] ; \quad \mathbf{z}^{s} \sim p(\mathbf{z}|\mathbf{E},\boldsymbol{\theta})
\end{split}
\end{equation}

\begin{table}[b]
\caption{Parameters to be estimated}
    \centering
    \label{tab:param}
 \begin{tabular}{| c | c |}
\hline
 Parameter & Description   \\
 \hline
$\mathbf{w}^{day}_c $ & Weight of week-based features for category $c$   \\
$\mathbf{w}^{time}_c $ & Weight of hour-based features for category $c$  \\
$\alpha $ & Influence matrix across categories  \\
$h $ & Bandwidth  \\
$\eta $ & Temporal decay  \\
\hline
\end{tabular}
\end{table}
\textbf{Gibbs sampling} Gibbs sampling is quite useful to obtain samples from the posterior of a random variable following discrete distribution. Its a  Markov Chain Monte Carlo (MCMC) approach which is used when the conditional probabilities can be computed tractably. Its an iterative approach where one starts with some random assignments to the latent variables. In each iteration a new sample for a latent variable is obtained from the conditional distribution while fixing all other latent variables to its previously assigned values. Unlike MCMC, all the latent variable values sampled from the conditional distribution is accepted.  After some iterations, the process converges and the samples are obtained from the posterior distribution over the latent variables. It requires us to compute the conditional probabilities $p(\mathbf{z}_i| \mathbf{z}_{-i}, \mathbf{E},\boldsymbol{\theta} )$. We compute the conditional probability as follows and is tractable -

\setlength{\belowdisplayskip}{1pt} \setlength{\belowdisplayshortskip}{1pt}
\setlength{\abovedisplayskip}{1pt} \setlength{\abovedisplayshortskip}{1pt}
\begin{equation}
    \begin{split}
 p(\mathbf{z}_i| \mathbf{z}_{-i}, \mathbf{E},\boldsymbol{\theta} ) &= \frac{ p (\mathbf{z}_i, \mathbf{z}_{-i} | \mathbf{E}, \boldsymbol{\theta}) } { p(\mathbf{z}_{-i}  | \mathbf{E}, \boldsymbol{\theta}) } \\ &=\frac{p(\mathbf{E} | \mathbf{z}, \boldsymbol{\theta}) p(\mathbf{z})}{\sum_{\mathbf{z}_i=1}^K  p(\mathbf{E} | \mathbf{z}_i, \mathbf{z}_{-i}, \boldsymbol{\theta}) p(\mathbf{z}_i,\mathbf{z}_{-i})  }
 \label{gibbs_samp_eqn}
    \end{split}
\end{equation}

Samples over all the latent variables are obtained by sampling from this distribution iteratively  and the samples after some burn-in iterations are used to compute the expected log joint likelihood.
 
\begin{algorithm}[ht!]
\caption{Algorithm for Semantic Annotation}
\label{alg:algorithm1}
\begin{algorithmic}[1]
\State {\textbf{Input:}} Check-in data $\mathbf{E}$ consisting of observed  and missing categories $\mathbf{z}$.
\State Initialize $\eta$, $h$, $\mathbf{w}^{day}_c $, $\mathbf{w}^{time}_c $ and $\alpha$ 
\Repeat
\State{\textbf{E-Step}}
\State Initialize $\mathbf{z}^{(1)}=(z_1^{1},..., z_k^{1})$ with some random categories 
\State Initialize number of iterations ($iter$) \& set of samples $S = \phi$
\For{$t\gets 1, iter$}
    \For{$i \gets 1,N_z$}
        \State Sample $ z' \sim p(z_i^{t}| \mathbf{z}_{-i}^{t}, \mathbf{E},\boldsymbol{\theta} )$ where $\mathbf{z}_{-i}^t $ refers to all variables in $\mathbf{z}^t$ except $i^{th}$ variable
        \State  $\mathbf{z}_{i}^{t} \gets z'$
    \EndFor
\State $S = S \bigcup \mathbf{z}^t$
\State $\mathbf{z}^{t + 1} \gets \mathbf{z}^{t}$
\EndFor
\State {\textbf{M-Step}}
\State Using the samples S to find $ \mathbb{E}_{p(\mathbf{z}|\mathbf{E},\boldsymbol{\theta})} [\log p(\mathbf{E},\mathbf{z} | \boldsymbol{\theta})]$
\State Update $\mathbf{w}^{day}_c , \mathbf{w}^{time}_c, \alpha \text{ maximizing } \mathbb{E}_{p(\mathbf{z}|\mathbf{E},\boldsymbol{\theta})} [\log p(\mathbf{E},\mathbf{z} | \boldsymbol{\theta})]$
\Until{convergence} \\
\Return{samples, $\mathbf{w}^{day}_c $, $\mathbf{w}^{time}_c $ and $\alpha$ }
\end{algorithmic}
\end{algorithm}

\subsubsection{\textbf{Maximization Step}}
In maximization-step, we will find the parameters associated with the model by maximizing the expected log joint likelihood with respect to the parameters using a gradient descent approach. We find $\nabla_{\boldsymbol{\theta}} LL$ with respect to all the parameters of our model. Table~\ref{tab:param} mentions the parameters to be estimated for our model. 
To find the derivatives with respect to different parameters, we will differentiate log-likelihood with respect to each parameter. We will differentiate parameters using the following -
\begin{equation}
   \nabla_{\boldsymbol{\theta}} \Tilde{L}(\boldsymbol{\theta}) =  \nabla_{\boldsymbol{\theta}}\frac{1}{N} \sum_{s=1}^S  [ \log p(\mathbf{E} | \mathbf{z}^{s}, \boldsymbol{\theta}) + \log p(\mathbf{z}^{s})] 
\end{equation}
 Algorithm \ref{alg:algorithm1} summarizes the methodology for our proposed model.

\section{Prediction}
\label{prediction_task}
In this section, we focus on the problem of \emph{prediction of future check-ins in the test interval}. For this, we need to predict \textit{timestamp}, \textit{latitude}, \textit{longitude} and \textit{category}.  We performed  this by using a modified Ogata's thinning algorithm \citep{paper23}, since the standard thinning algorithm is used to sample points from continuous space whereas location-based social networks deal with discrete locations. The algorithm uses an idea similar to rejection sampling and finds the next user location based on his intensity function. Human dynamics generally involve movement around close-by regions and hence, we used Gaussian centered around last check-in as  proposal distribution to sample locations in the rejection sampling.
We consider a discrete set of locations $L$ visited by user. We sample $(x_{curr}, y_{curr})$ from a Gaussian centered at the previous coordinate $(x_{prev}, y_{prev})$. Then we find the nearest discrete location $(x_{disc}, y_{disc})$ associated with the sampled point $(x_{curr}, y_{curr})$ from $L$. This discrete location is associated with the $venue-id$. In order to sample category, we sample from multinoulli distribution with the parameters of distribution being proportionate to the the intensity for respective categories. In order to find the next point, we simulate a homogenous point process with intensity $\lambda^{*}$ which is an upper bound of intensity $\lambda(t,x,y)$ in the desired interval. For temporal Hawkes process with monotonically decreasing kernel in the interval $[t_i, t_{i+1})$, we observe that $\lambda^{*}$ can be found as $\lambda(t_i)$.
Although both temporal and spatial kernels are monotonically decreasing but the past influence arising from it may not be always monotonically decreasing. We can approximate $\lambda^{*}$ with $\lambda(t,x,y)$. Also, we accept a point by considering time and latitude-longitude pair jointly with a probability of $\dfrac{\lambda(t,x,y)}{\lambda^{*}}$. This process is repeated to predict the check-ins. Algorithm \ref{alg:algorithm2} presents the detailed steps for prediction of check-ins.
\begin{algorithm}[t]
\caption{Algorithm for Prediction}
\label{alg:algorithm2}
\begin{algorithmic}[1]
\State {\textbf{Input:}} Last check-in information $(u_n, t_n, x_n, y_n, c_n)$, Past check-ins $\mathbf{E}$, $\mathbf{z}$, $\eta$, $h$,  $\mathbf{w}^{day}_c$ , $\mathbf{w}^{time}_c$ , $\alpha$ \text{ and } $T $
\Repeat
\State Initialize $t = t_n$, $x = x_n$, $y = y_n$ , $u = u_n$, $c = c_n$, $S = \phi$
\Repeat
    \State Calculate $\lambda^{*} \gets \sum_{c=1}^K\lambda^u_c(t,x,y)$
    \State Sample $(x_{curr}, y_{curr}) \sim \mathcal{N}((x,y),h)$
    \State Find discrete location $(x_{new}, y_{new})$ closest to $(x_{curr}, y_{curr})$
    \State Sample $s \sim \exp(1/\lambda^*)$
    \State $t_{new} \gets t + s$
    \State Calculate $ \lambda_{new} \gets  \sum_{c=1}^K \lambda^u_c(t_{new}, x_{new}, y_{new}) $
    \State Sample $u \sim U[0,1]$
    \If{$u < \lambda_{new}/\lambda^*$}
        \State prob = []
        \For{$k\gets 1, C$}
            \State $\lambda_{k} \gets \lambda^u_k(t_{new}, x_{new}, y_{new})$
            \State $prob[k] \xleftarrow{} \lambda_{k} $
        \EndFor
        \State $c' \sim Cat(prob)$
        \State $S \gets S \bigcup (t_{new}, x_{new},y_{new},u,c') $
    \EndIf
    \State $t \gets t_{new}, x \gets x{new}, y \gets y_{new}$
\Until{$t > T$} 
\Until{checkins for each user is predicted} \\
\Return{$S$}
\end{algorithmic}
\end{algorithm}

\section{Experiments}
In this section, we will discuss about our experiments and results for the proposed approach. We perform the tasks of semantic annotation of missing categories and location adoption modelling and discuss each in the following sections. However, the main focus of this work is encompassed around semantic annotation of missing categories. 
The goal of performing location adoption dynamics is merely for the qualitative assessment of the proposed approach.
\subsection{Dataset Description}
We have conducted our experiments on two widely used LBSN datasets: Foursquare \citep{paper22} and Gowalla \citep{paper2}. Gowalla contains check-in data ranging from January 2009 to August 2010, and Foursquare includes the check-in data ranging from December 2009 to June 2013. Each check-in includes user-id, venue-id, timestamp, latitude, longitude and category. Foursquare has around 400 fine-grained categories in their dataset. We have replaced them with broader categories using the category tree structure mentioned on the Foursquare website \footnote[1]{https://developer.foursquare.com/docs/build-with-foursquare/categories/}. Through this process we have converged to eight categories for Gowalla and nine categories for Foursquare. A similar process has been followed by \citep{paper3} as well. This helps to get a wider perspective over category-category influence and hence aids in meaningful interpretation. We have created a
representative subset of both datasets to conduct all the experiments. All the check-ins are considered during a twelve week period in order to capture the sequential nature of events. Out of this subset, last four weeks data is considered as test data for prediction of check-ins.  In the remaining data, we have randomly removed the category associated with a check-in to add missing categories in the dataset. To prove robustnes of our experiments for the task of semantic annotation of missing data, we have created datasets with both 10\% and 20\% missing categories. 

In our experimental set-up, we have considered bandwidth ($h$) and temporal decay ($\eta$) to be hyper-parameters. The values of these hyper-parameters have been set through grid search. Also, we have used shared parameters across all the users. Different regularization constants have been used for $\mathbf{w}_{day}$,  $\mathbf{w}_{time}$ and $\alpha$. Moreover, the weights for week-based features ($\mathbf{w}_{day}$), hour-based features ($\mathbf{w}_{time}$) and influence matrix ($\alpha$) are randomly initialized. Moreover, to improve the generalization ability of the proposed model, we have used l2-regularization over all parameters. For Gibbs Sampling, we have set total iterations to be 1000 and burning iterations to be 750. In each such iteration, a category is sampled for each missing event using the conditional probabilities for each category  \eqref{gibbs_samp_eqn}  discussed in Section \ref{param_est_estep}. The intensity values are used for calculating conditional probabilities during Gibbs sampling. This is repeated for all missing events in an iteration.   After getting samples from Gibbs sampling procedure, we select every third sample to get independent samples. The expectation-maximization steps are repeated till we achieve relative convergence of $10^{-10}$. 

\begin{figure}[t] 
    \centering
        \includegraphics[width=0.33\textwidth, height=0.33
        \textwidth,keepaspectratio]{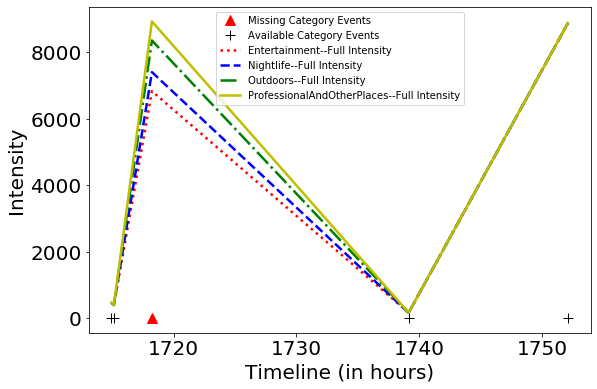}
    \caption{Plot representing intensity of top four categories for the missing event with actual category as Professional from the Gowalla dataset. The intensity values are obtained during the conditional probability calculation in the Gibbs sampling step for the  missing event.}
    \label{fig:intensity_plots}
\end{figure}

\begin{figure}
    \centering
   \includegraphics[width=0.38\textwidth, height=0.38\textwidth,keepaspectratio]{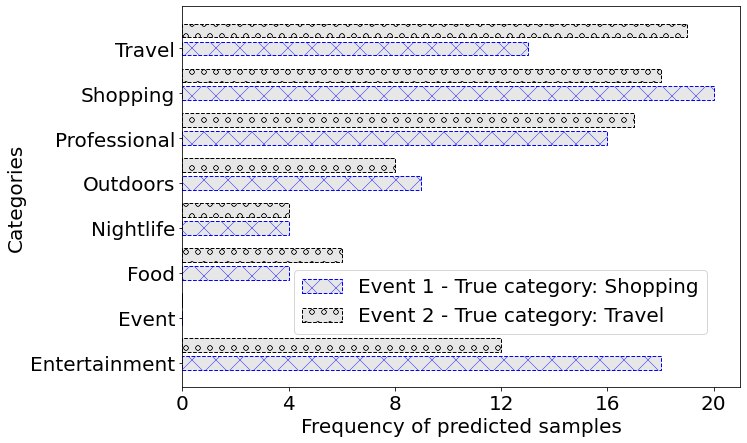}
    \caption{An example of sampled categories for two missing events which we get as output from our model. Each sample consists of distribution of categories for a missing event. }

    \label{fig:prediction_samples}
\end{figure}

In Figure \ref{fig:intensity_plots}, we show the intensity function values for top four  categories for a missing event with actual category \textit{Professional} from the last iteration of Gibbs sampling step. We can observe that the intensity function value for \textit{Professional} is the highest, and consequently we sample the category \textit{Professional}. The output of our model is a set of samples corresponding to all the missing events which we will refer as \textit{sampled categories}. Sampled categories represent the posterior distribution over the categories and are obtained as an output of Gibbs sampling. Each sample consists of distribution of categories for the missing event. An example of sampled categories for two missing events is depicted in Figure~\ref{fig:prediction_samples}. 


\subsection{Experiments and results for semantic annotation of missing data}
We will now describe the experimental details for the task of semantic annotation of missing categories using proposed model on both datasets. As already stated, we are performing missing category prediction per event. To the best of our knowledge this is the first work for semantic annotation in LBSNs where we are predicting category for each event. However, there exist approaches where semantic annotation is performed for venue. Hence, we will adapt our model to predict the venues so that we can compare with an existing baseline. We will evaluate the performance of our model on both venue and event level.  
We have decomposed the assessment of our model for the task of semantic annotation across these two dimensions in the following way: 

\begin{table*}[ht]
    \centering
    \caption{Results for Venue-centric Precision, Recall and F-score}
    \label{tab:venuebasedresult}

\begin{tabular}{cccccccc|cccccc}
           & ~         & \multicolumn{6}{c|}{Missing Categories 10\%}                                          & \multicolumn{6}{c}{Missing Categories 20\%}                                           \\ 
\hline
           &           & \multicolumn{2}{c}{SAP \citep{paper3}} & \multicolumn{2}{c}{PPE \citep{paper17}} & \multicolumn{2}{c|}{HAP-SAP} & \multicolumn{2}{c}{SAP \citep{paper3}} & \multicolumn{2}{c}{PPE \citep{paper17}} & \multicolumn{2}{c}{HAP-SAP}  \\ 
\hline
Datasets   & Measure   & Micro & Macro           & Micro & Macro           & Micro & Macro                     & Micro & Macro           & Micro & Macro           & Micro & Macro                     \\ 
\hline
Gowalla    & Precision & 0.1   & 0.025           & 0.172 & 0.158           & 0.166 & 0.138                     & 0.102 & 0.096           & 0.107 & 0.106           & 0.141 & 0.130                     \\
           & Recall    & 0.088 & 0.087           & 0.147 & 0.147           & 0.441 & 0.370                     & 0.177 & 0.163           & 0.215 & 0.221           & 0.531 & 0.461                     \\
           & F1-Score  & 0.093 & 0.039           & 0.158 & 0.117           & 0.241 & 0.196                     & 0.129 & 0.069           & 0.143 & 0.08            & 0.223 & 0.196                     \\ 
\hline
Foursquare & Precision & 0.208 & 0.170           & 0.17  & 0.04            & 0.119 & 0.12                     & 0.200 & 0.118           & 0.15  & 0.16            & 0.122 & 0.119                     \\
           & Recall    & 0.333 & 0.277           & 0.26  & 0.11            & 0.866 & 0.611                     & 0.351 & 0.259           & 0.16  & 0.13            & 0.567 & 0.580                     \\
           & F1-Score  & 0.256 & 0.207           & 0.21  & 0.05            & 0.209 & 0.199                     & 0.254 & 0.148           & 0.15  & 0.1             & 0.201 & 0.191                    
\end{tabular}
\end{table*}

\begin{table}
    \centering
\caption{Venue-centric Accuracy (in \%)}
    \label{tab:venuebasedaccuracy}
    \tabcolsep=0.1cm
   \begin{tabular}{ccccccc}
Dataset                     & \begin{tabular}[c]{@{}c@{}}Missing \\Categories\end{tabular} & SAP \citep{paper3}                  & PPE \citep{paper17}                  & \multicolumn{3}{c}{HAP-SAP}  \\ 
\hline
\multicolumn{1}{l}{}        & \multicolumn{1}{l}{}                                         & \multicolumn{1}{l}{} & \multicolumn{1}{l}{} & Top-1 & Top-2 & All               \\ 
\hline
Gowalla & 10\% &8.82  & 14.7  & 26.47 & 41.17 & 44.11             \\
                            & 20\%                                                         & 17.17                & 21.51                & 15.18 & 26.58 & 53.16             \\ 
\hline
Foursquare & 10\%                                                         & 33.33                & 26.26                & 20    & 26.66    & 86.66             \\
                            & 20\%                                                         & 35.13                & 16.21                & 21.62 & 32.43 & 56.75            
\end{tabular}
    
\end{table}

\subsubsection{\textbf{Venue-centric experiments}} \hfill\\
We have compared our proposed approach     (\textbf{HAP-SAP}) with the following baselines:
\begin{itemize}
\item \textbf{Semantic Annotation of Places (SAP) \citep{paper3}:} This work uses various features like population features (e.g., number of unique visitors) and temporal features (e.g., distribution of check-in time) as semantic descriptions of specific places. Then it develops a network of related places using random walk and restart where they derive the probability for a specific tag being labeled to a place from its similar places. This label probability is used along with population and temporal features, to feed to the binary SVM classifiers for each category to predict associated categories with the venues in SAP algorithm.
\item \textbf{Predictive Place Embedding (PPE) \citep{paper17}:} This is a state-of-the-art POI tag annotation method. In this work, authors have proposed a graph embedding method to learn POI embedding using a user-tag  and POI-temporal bipartite graph. The learnt POI embedding vectors are used as input to multi class SVM classifier. For learning the parameters, we use edge sampling and negative sampling.
\end{itemize}

\textbf{Evaluation:} We have used \textit{precision, recall, F1-score and accuracy} scores at micro and macro levels for evaluating our model. Now we will explain how we are adapting our model to calculate venue based accuracy. 

In order to compare our experiments with baselines, we introduce a methodology to calculate the venue-centric scores for our test samples. Under our model set-up, there may be some venues whose categories are never observed in training set since we have removed the categories randomly. Such venues will be referred as \textit{unseen venues} and the rest are referred as \textit{seen venues}. We will report results for unseen venues only since categories for seen venues are already available in training dataset. Since one venue may be associated with more than one event, we have grouped together the sampled categories associated with each venue over all events to calculate venue based accuracy.   After this process, each venue has a set of categories aggregated from the sampled categories of the events. Evaluations will be performed on this set of categories. 

For all the evaluations, we classify a prediction to be correct in terms of number of hits. We define a hit to be 1 if we predict at least one category correctly and 0 otherwise. Thus, accuracy is defined as 
 \begin{equation*}
Accuracy = \dfrac{\#hits}{|S_{test}|}
 \end{equation*}
where \#hits represents number of times when we predict atleast one correct category and $|S_{test}|$ represents number of test samples. 
 Since our method is based on distribution of samples, we have calculated accuracy by considering Top-k categories. For this, we rank the sampled categories based on frequency of samples and use k most frequent categories, which is referred as Top-k accuracy. And 'All' considers all the predicted categories of the samples after prediction. For the calculation of precision, recall and F1-score, we have considered the categories associated with all the samples across all events associated with each venue. 
 
The previous works, SAP and PPE, are based on multiple binary classifiers (one versus rest), so more than one category can be predicted for each venue. In such a setting, we have computed macro score by considering the  metrics per  categories and then taking the average. Micro scores are calculated using the sum of all true positives(TP), false positives(FP), and false negatives(FN) over all labels. These values are then used to find the precision, recall and accuracy using respective formulae. Considering number of categories to be $K$, the formulae for macro and micro scores can be written as follows - 
\begin{equation*}
\begin{split}
    & Micro\text{-}Precision = \frac{\sum_{i=1}^K TP_i}{\sum_{i=1}^K TP_i + \sum_{i=1}^K FP_i} \\
    & Micro\text{-}Recall = \frac{ \sum_{i=1}^K TP_i }{\sum_{i=1}^K TP_i + \sum_{i=1}^K FN_i} \\
    & Micro\text{-}F1 Score = \frac{2*Micro\text{-}Precision*Micro\text{-}Recall}{Micro\text{-}Precision + Micro\text{-}Recall} \\
    & Macro\text{-}Precision = \frac{\sum_{i=1}^K Precision}{K} \\
    & Macro\text{-}Recall = \frac{\sum_{i=1}^K Recall}{K} \\
    & Macro\text{-}F1 Score = \frac{2*Macro\text{-}Precision*Macro\text{-}Recall} {Macro\text{-}Precision + Macro\text{-}Recall}
\end{split}
\end{equation*}

 The venue-based results for precision, recall and F-score are reported in Table~\ref{tab:venuebasedresult}. We can observe from both the tables that recall scores for our model always outperform both the baselines substantially for both the datasets. Although we don't observe a significant increment in precision for Foursquare dataset, however we get better F1-score for most of the settings. For Gowalla dataset, we get better precision for 20\% missingness than both the models. However, for 10\% missingness, it's very close to PPE but outperforms SAP. F1-scores are better for both 10\% and 20\% missingness for Gowalla dataset. 

Accuracy results for the venues are reported in  Table~\ref{tab:venuebasedaccuracy}. The results suggest that our results are better than both the baselines. We can also see that merely Top-2 accuracy results for our model are better than baselines in all cases. Also, if we consider all the categories, our results are far better than both SAP and PPE. We can also observe that Top-1 and Top-2 accuracy, although being better than baselines, drops for 20\% missing categories. A possible explanation for this could be due to increase the sampling dimensionality.  

\begin{table}
\centering
\caption{Event-centric accuracy (in \%)}
\label{tab:eventbasedaccuracy}
 \tabcolsep=0.1cm
\begin{tabular}{cccccc}
Dataset    & \begin{tabular}[c]{@{}c@{}}Missing\\ Categories \end{tabular} & \begin{tabular}[c]{@{}c@{}}SVM\\Features \end{tabular} & \multicolumn{3}{c}{HAP-SAP}  \\ 
\hline
           &                                                                       &                                                        & Top-1 & Top-2 & Top-3             \\ 
\hline
Gowalla    & 10\%                                                                  & 27.53                                                  & 21.01 & 31.15 & 36.23             \\
           & 20\%                                                                  & 31.88                                                  & 14.13 & 24.27 & 34.07             \\ 
\hline
Foursquare & 10\%                                                                  & 16.98                                                  & 18.86 & 33.96 & 45.91             \\
           & 20\%                                                                  & 17.2                                                   & 16.30 & 26.01 & 34.38            
\end{tabular}
\end{table}
\subsubsection{\textbf{Event-centric experiments}} \hfill\\
We have compared our proposed method with an SVM based approach where we use temporal features similar to the one used in HAP-SAP. Here, we include distribution for check-in day and check-in time at event level also in addition to venue level, making it a more competitive baseline. 
On combining these features, we feed this to binary SVM based classifiers for each category (one vs. rest). We will refer this approach as \textit{SVMFeatures}. This method can predict multiple categories for a check-in event. 

 For evaluation, we have used accuracy of the category assigned to each event. Since output of HAP-SAP consists of distribution of samples for each missing event, we calculate accuracy by considering Top-k samples. This is termed as \textit{Acc@k}. Similar to venue-centric evaluation, we rank categories of the predicted samples based on frequency. Acc@k refers to the Top-k categories from this ranked list. We have reported event-based accuracy in Table~\ref{tab:eventbasedaccuracy}. The table illustrates that HAP-SAP outperforms SVMFeatures for both datasets. Similar to venue based accuracy, we can observe here also that Top-2 accuracy is better than baseline in most of the cases. Also, results for 10\% missing categories are better than 20\% missing categories.   Although SVMFeatures used features based on venue as well as event, our model performs better than the baseline. This shows the relevance of considering sequential information over event based features. 
 

\subsection{Experiments and results for prediction}
The additional advantage of our model over previous work is that we can model location adoption in a location-based social network framework. The goal of performing location adoption dynamics also helps for qualitative assessment of correct inference of missing categories. We conduct experiments to demonstrate that the categories inferred by our model could improve the location prediction task.   We can predict future check-ins using Algorithm \ref{alg:algorithm2} in  Section \ref{prediction_task}. We have performed prediction using lookahead of one where we predict one timestep ahead by considering the actual historical set of events. For prediction of future events given a set of check-ins with missing categories, we compare our work with the following baselines which are often considered as general accepted ways of missing data imputation for categorical data:
\begin{itemize}
    \item \textbf{RandomCat:} In this method, we assign random categories to the missing events and predict the future check-ins. 
    \item \textbf{RemoveMissing:} This method removes the check-ins where the categories are missing. Thereafter, we try to predict the future events.
\end{itemize}
We evaluate our prediction model by calculating Root Mean Squared Error for the timestamp of the predicted events.
Root Mean Squared Error (RMSE) is defined as - 
\[ RMSE = \sqrt{\dfrac{1}{n} \sum_{i=1}^{N} (y_{i} - \hat{y_{i}})^2}
\]


\begin{figure}
    \centering
    \vspace{-5mm}
    \includegraphics[width=0.40\textwidth, height=0.40\textwidth,keepaspectratio]{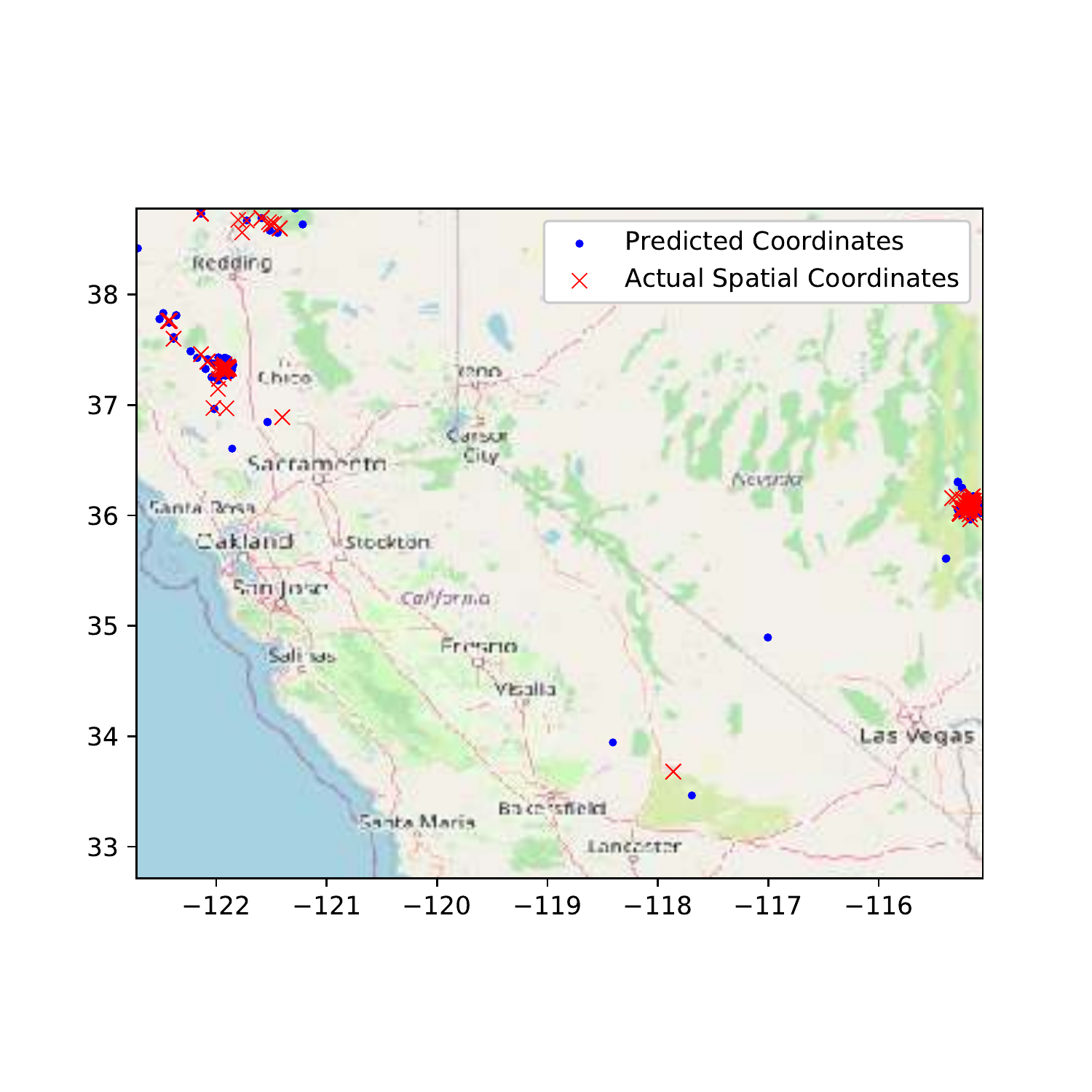}
    \vspace{-13mm}
    \caption{Analysis of Spatial Predictive Performance for Diffusion modelling for Gowalla Dataset }
    \label{fig:spatial_predictions_all}
    \vspace*{-2mm}
\end{figure}

\begin{table}[]
\centering
\begin{tabular}{ccc}
Model           & Gowalla & Foursquare  \\ 
\hline
HAP-SAP    & 1.210  & 0.717       \\
Random Samples  & 1.288   & 0.805      \\
Removed Samples & 3.024  & 3.007    
\end{tabular}
\vspace{5mm}
\caption{Temporal Predictive Performance (in hours)}
\vspace{-5mm}
\label{tab:predictionresult}
\end{table}

Table~\ref{tab:predictionresult} provides the results of prediction. A comparison of mean squared error achieved by our model and the proposed baselines suggest that our model performs better than the discussed baseline approaches. In fact, our results reflect significant difference between the values of our model as compared to the model where we remove events with missing category.
Also, we analyze the spatial plots for the predicted events. For this we plot the predicted latitude-longitude pairs with the predicted ones. Figure~\ref{fig:spatial_predictions_all} depicts the spatial plots for both the datasets. We can observe that the spatial coordinates predicted for Gowalla datset is very close to the actual coordinates, indicating the efficiency of spatial predictions as well. Therefore, the prediction results corroborate the relevance of missing category imputation in location-based social networks in which sequence information is pertinent.

\subsection{Analysis}
\begin{figure}[] 
    \centering
    \subfloat[]{
  \includegraphics[width=0.40\textwidth, height=0.40\textwidth,keepaspectratio]{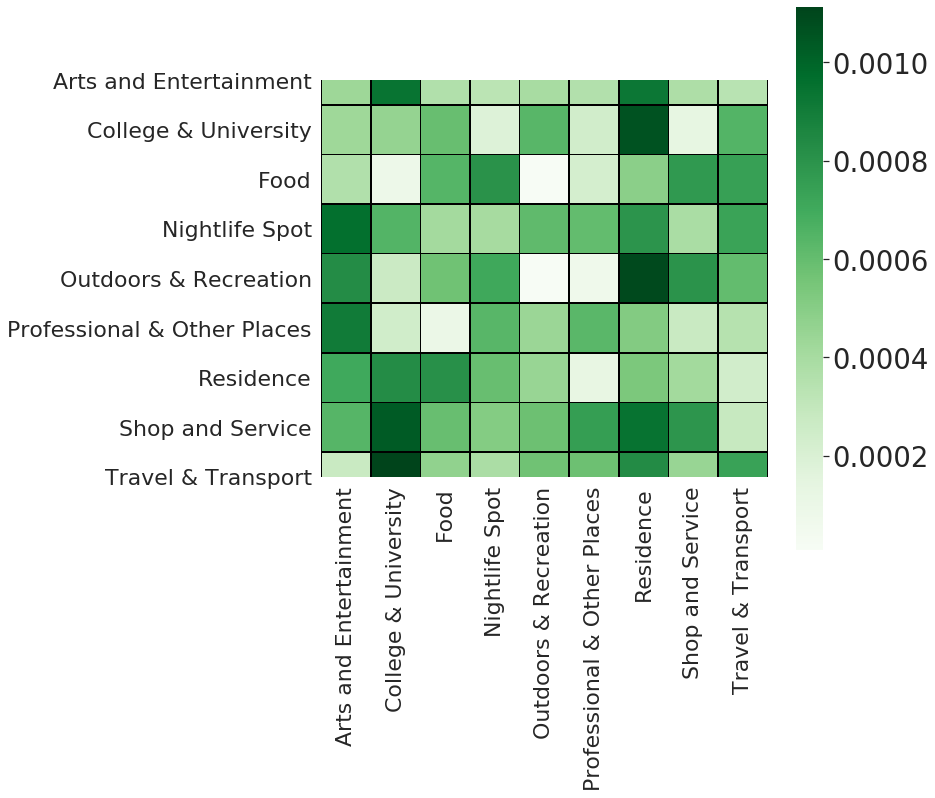}
  \label{foursquare_alpha}
  }
  
\subfloat[]{
  \includegraphics[width=0.30\textwidth, height=0.30\textwidth,keepaspectratio]{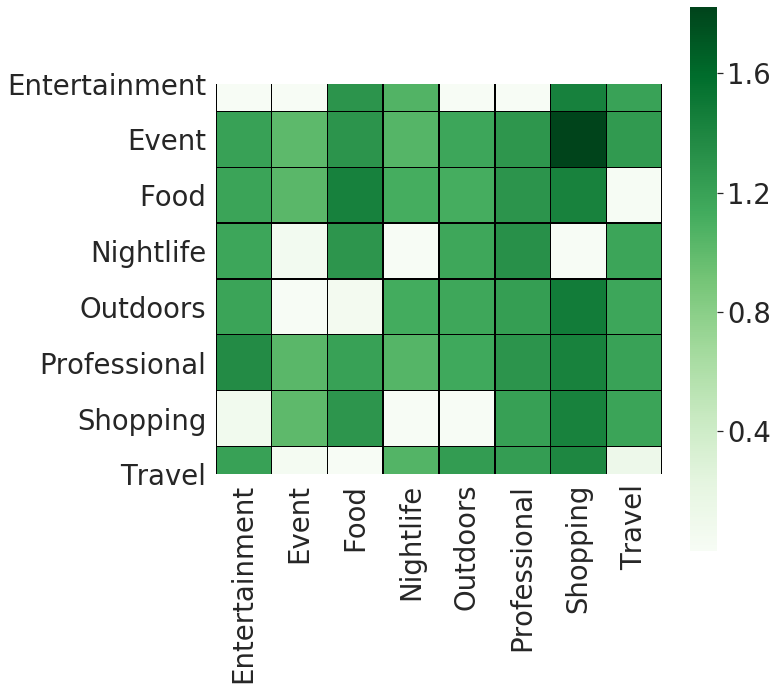}
  \label{gowalla_alpha}
  }
    \caption{Influence matrix for Figure~\ref{foursquare_alpha}) Foursquare and Figure~\ref{gowalla_alpha}) Gowalla datset}
    \label{fig:influencematrix}
\end{figure}

We analyze the values of influence matrix learned by our model for Foursquare and Gowalla dataset with 10\% missing categories. The Gowalla dataset consists of 8 categories. Hence, the dimension of influence matrix is $8 \times 8$. It captures influence of one category over another. $\alpha_{ij}$ represents the causation from $j^{th}$ category to $i^{th}$ category. As can be observed from Figure \ref{fig:influencematrix}, there is a high influence from \textit{Travel} category to other categories, which is very intuitive. Similarly, there is high influence from \textit{Professional} to \textit{Food} and \textit{Nightlife}. On the contrary, we can observe a low influence from \textit{Nightlife} to \textit{Shopping}. Among similar category transitions, we can see low influence from \textit{Nightlife} to \textit{Nightlife} because it is less likely to go from one venue to another for nightlife. Whereas we observe high influence for \textit{Shopping} to \textit{Shopping} and \textit{Food} to \textit{Food} because it is very common to move from one shopping or food venue to another. Hence, influence matrix captures transition from one category to another very well. Similarly, we can analyze influence matrix for Foursquare as well. The 
the dimension of influence matrix for Foursquare is $9 \times 9$. We can observe high values from \textit{Travel and Transport} to \textit{College and University}, \textit{Outdoors and Recreation} to \textit{Residence} and \textit{College and University} to \textit{Residence} which is pretty intuitive. Moreover, the light colored cells represents less tendency to move from \textit{Food} to \textit{Outdoors and Recreation} and \textit{Outdoors and Recreation} to \textit{Outdoors and Recreation}. In this way, such an analysis can be useful to understand the causation relationships.

\section{Conclusion}
In this paper, we propose a latent multivariate spatio-temporal Hawkes process to perform semantic annotation in location-based social networks known as HAP-SAP. 
We use Hawkes process which assumes a  self-triggering property that the occurrence of an event will influence future events. Such model can capture inhomogeneous inter-event times and causal correlations, which are important considerations for human dynamics. 
Since location-based social networks are hugely affected by geographical influence, it is imperative to use spatial information within the framework of Hawkes process using a spatial kernel. This allows to build an extensive model to incorporate temporal factors, geographical influence and user-interests. 
Each check-in category is considered as mark associated with the event. We model the missing categories associated with the event as latent marks. We employ expectation-maximization procedure to infer the missing categories and learn model parameters. We obtain samples from the posterior distribution of latent categories using Gibbs sampling and use them in computing expectation. Consequently, we associate a category to each check-in. Then we use the learnt parameters to understand mobility dynamics and predict future check-ins. Our experiments on real world datasets proves the effectiveness of our model and reinforces the relevance of missing category imputation for an event in location-based social networks.

\bibliographystyle{ACM-Reference-Format}
\bibliography{sample-sigconf}


\begin{thebibliography}{22}


\ifx \showCODEN    \undefined \def \showCODEN     #1{\unskip}     \fi
\ifx \showDOI      \undefined \def \showDOI       #1{#1}\fi
\ifx \showISBNx    \undefined \def \showISBNx     #1{\unskip}     \fi
\ifx \showISBNxiii \undefined \def \showISBNxiii  #1{\unskip}     \fi
\ifx \showISSN     \undefined \def \showISSN      #1{\unskip}     \fi
\ifx \showLCCN     \undefined \def \showLCCN      #1{\unskip}     \fi
\ifx \shownote     \undefined \def \shownote      #1{#1}          \fi
\ifx \showarticletitle \undefined \def \showarticletitle #1{#1}   \fi
\ifx \showURL      \undefined \def \showURL       {\relax}        \fi
\providecommand\bibfield[2]{#2}
\providecommand\bibinfo[2]{#2}
\providecommand\natexlab[1]{#1}
\providecommand\showeprint[2][]{arXiv:#2}

\bibitem[\protect\citeauthoryear{Cho, Myers, and Leskovec}{Cho
  et~al\mbox{.}}{2011}]%
        {paper2}
\bibfield{author}{\bibinfo{person}{Eunjoon Cho}, \bibinfo{person}{Seth~A
  Myers}, {and} \bibinfo{person}{Jure Leskovec}.}
  \bibinfo{year}{2011}\natexlab{}.
\newblock \showarticletitle{Friendship and mobility: user movement in
  location-based social networks}. In \bibinfo{booktitle}{\emph{Proceedings of
  the 17th ACM SIGKDD international conference on Knowledge discovery and data
  mining}}. \bibinfo{pages}{1082--1090}.
\newblock


\bibitem[\protect\citeauthoryear{Cho, Galstyan, Brantingham, and Tita}{Cho
  et~al\mbox{.}}{2013}]%
        {paper36}
\bibfield{author}{\bibinfo{person}{Yoon-Sik Cho}, \bibinfo{person}{Aram
  Galstyan}, \bibinfo{person}{P~Jeffrey Brantingham}, {and}
  \bibinfo{person}{George Tita}.} \bibinfo{year}{2013}\natexlab{}.
\newblock \showarticletitle{Latent self-exciting point process model for
  spatial-temporal networks}.
\newblock \bibinfo{journal}{\emph{arXiv preprint arXiv:1302.2671}}
  (\bibinfo{year}{2013}).
\newblock


\bibitem[\protect\citeauthoryear{Diggle, Rowlingson, and Su}{Diggle
  et~al\mbox{.}}{2005}]%
        {paper18}
\bibfield{author}{\bibinfo{person}{Peter Diggle}, \bibinfo{person}{Barry
  Rowlingson}, {and} \bibinfo{person}{Ting-li Su}.}
  \bibinfo{year}{2005}\natexlab{}.
\newblock \showarticletitle{Point process methodology for on-line
  spatio-temporal disease surveillance}.
\newblock \bibinfo{journal}{\emph{Environmetrics: The official journal of the
  International Environmetrics Society}} \bibinfo{volume}{16},
  \bibinfo{number}{5} (\bibinfo{year}{2005}), \bibinfo{pages}{423--434}.
\newblock


\bibitem[\protect\citeauthoryear{Hainzl, Steacy, and Marsan}{Hainzl
  et~al\mbox{.}}{2010}]%
        {paper20}
\bibfield{author}{\bibinfo{person}{Sebastian Hainzl}, \bibinfo{person}{D
  Steacy}, {and} \bibinfo{person}{S Marsan}.} \bibinfo{year}{2010}\natexlab{}.
\newblock \showarticletitle{Seismicity models based on Coulomb stress
  calculations}.
\newblock \bibinfo{journal}{\emph{Community Online Resource for Statistical
  Seismicity Analysis}} (\bibinfo{year}{2010}).
\newblock


\bibitem[\protect\citeauthoryear{Hawkes}{Hawkes}{1971}]%
        {paper24}
\bibfield{author}{\bibinfo{person}{Alan~G Hawkes}.}
  \bibinfo{year}{1971}\natexlab{}.
\newblock \showarticletitle{Spectra of some self-exciting and mutually exciting
  point processes}.
\newblock \bibinfo{journal}{\emph{Biometrika}} \bibinfo{volume}{58},
  \bibinfo{number}{1} (\bibinfo{year}{1971}), \bibinfo{pages}{83--90}.
\newblock


\bibitem[\protect\citeauthoryear{He, Yin, Chen, Zhou, Sadiq, and Luo}{He
  et~al\mbox{.}}{2016}]%
        {paper16}
\bibfield{author}{\bibinfo{person}{Tieke He}, \bibinfo{person}{Hongzhi Yin},
  \bibinfo{person}{Zhenyu Chen}, \bibinfo{person}{Xiaofang Zhou},
  \bibinfo{person}{Shazia Sadiq}, {and} \bibinfo{person}{Bin Luo}.}
  \bibinfo{year}{2016}\natexlab{}.
\newblock \showarticletitle{A spatial-temporal topic model for the semantic
  annotation of POIs in LBSNs}.
\newblock \bibinfo{journal}{\emph{ACM Transactions on Intelligent Systems and
  Technology (TIST)}} \bibinfo{volume}{8}, \bibinfo{number}{1}
  (\bibinfo{year}{2016}), \bibinfo{pages}{1--24}.
\newblock


\bibitem[\protect\citeauthoryear{Le}{Le}{2017}]%
        {paper39}
\bibfield{author}{\bibinfo{person}{Triet~M Le}.}
  \bibinfo{year}{2017}\natexlab{}.
\newblock \showarticletitle{A multivariate hawkes process with gaps in
  observations}.
\newblock \bibinfo{journal}{\emph{IEEE Transactions on Information Theory}}
  \bibinfo{volume}{64}, \bibinfo{number}{3} (\bibinfo{year}{2017}),
  \bibinfo{pages}{1800--1811}.
\newblock


\bibitem[\protect\citeauthoryear{Lewis and Shedler}{Lewis and Shedler}{1979}]%
        {paper23}
\bibfield{author}{\bibinfo{person}{PA~W Lewis} {and} \bibinfo{person}{Gerald~S
  Shedler}.} \bibinfo{year}{1979}\natexlab{}.
\newblock \showarticletitle{Simulation of nonhomogeneous Poisson processes by
  thinning}.
\newblock \bibinfo{journal}{\emph{Naval research logistics quarterly}}
  \bibinfo{volume}{26}, \bibinfo{number}{3} (\bibinfo{year}{1979}),
  \bibinfo{pages}{403--413}.
\newblock


\bibitem[\protect\citeauthoryear{Li, Zhao, Zhang, Yuan, and Wang}{Li
  et~al\mbox{.}}{2020}]%
        {paper32}
\bibfield{author}{\bibinfo{person}{Yanhui Li}, \bibinfo{person}{Xiangguo Zhao},
  \bibinfo{person}{Zhen Zhang}, \bibinfo{person}{Ye Yuan}, {and}
  \bibinfo{person}{Guoren Wang}.} \bibinfo{year}{2020}\natexlab{}.
\newblock \showarticletitle{Annotating semantic tags of locations in
  location-based social networks}.
\newblock \bibinfo{journal}{\emph{GeoInformatica}} \bibinfo{volume}{24},
  \bibinfo{number}{1} (\bibinfo{year}{2020}), \bibinfo{pages}{133--152}.
\newblock


\bibitem[\protect\citeauthoryear{Linderman, Wang, and Blei}{Linderman
  et~al\mbox{.}}{2017}]%
        {paper35}
\bibfield{author}{\bibinfo{person}{Scott~W Linderman}, \bibinfo{person}{Yixin
  Wang}, {and} \bibinfo{person}{David~M Blei}.}
  \bibinfo{year}{2017}\natexlab{}.
\newblock \showarticletitle{Bayesian inference for latent Hawkes processes}.
\newblock \bibinfo{journal}{\emph{Advances in Neural Information Processing
  Systems}} (\bibinfo{year}{2017}).
\newblock


\bibitem[\protect\citeauthoryear{Liniger}{Liniger}{2009}]%
        {paper25}
\bibfield{author}{\bibinfo{person}{Thomas~Josef Liniger}.}
  \bibinfo{year}{2009}\natexlab{}.
\newblock \emph{\bibinfo{title}{Multivariate hawkes processes}}.
\newblock \bibinfo{thesistype}{Ph.D. Dissertation}. \bibinfo{school}{ETH
  Zurich}.
\newblock


\bibitem[\protect\citeauthoryear{Mei, Qin, and Eisner}{Mei
  et~al\mbox{.}}{2019}]%
        {paper38}
\bibfield{author}{\bibinfo{person}{Hongyuan Mei}, \bibinfo{person}{Guanghui
  Qin}, {and} \bibinfo{person}{Jason Eisner}.} \bibinfo{year}{2019}\natexlab{}.
\newblock \showarticletitle{Imputing missing events in continuous-time event
  streams}.
\newblock \bibinfo{journal}{\emph{arXiv preprint arXiv:1905.05570}}
  (\bibinfo{year}{2019}).
\newblock


\bibitem[\protect\citeauthoryear{Mohler, Short, Brantingham, Schoenberg, and
  Tita}{Mohler et~al\mbox{.}}{2011}]%
        {paper19}
\bibfield{author}{\bibinfo{person}{George~O Mohler}, \bibinfo{person}{Martin~B
  Short}, \bibinfo{person}{P~Jeffrey Brantingham},
  \bibinfo{person}{Frederic~Paik Schoenberg}, {and} \bibinfo{person}{George~E
  Tita}.} \bibinfo{year}{2011}\natexlab{}.
\newblock \showarticletitle{Self-exciting point process modeling of crime}.
\newblock \bibinfo{journal}{\emph{J. Amer. Statist. Assoc.}}
  \bibinfo{volume}{106}, \bibinfo{number}{493} (\bibinfo{year}{2011}),
  \bibinfo{pages}{100--108}.
\newblock


\bibitem[\protect\citeauthoryear{Rasmussen}{Rasmussen}{2013}]%
        {paper31}
\bibfield{author}{\bibinfo{person}{Jakob~Gulddahl Rasmussen}.}
  \bibinfo{year}{2013}\natexlab{}.
\newblock \showarticletitle{Bayesian inference for Hawkes processes}.
\newblock \bibinfo{journal}{\emph{Methodology and Computing in Applied
  Probability}} \bibinfo{volume}{15}, \bibinfo{number}{3}
  (\bibinfo{year}{2013}), \bibinfo{pages}{623--642}.
\newblock


\bibitem[\protect\citeauthoryear{Reinhart et~al\mbox{.}}{Reinhart
  et~al\mbox{.}}{2018}]%
        {paper26}
\bibfield{author}{\bibinfo{person}{Alex Reinhart} {et~al\mbox{.}}}
  \bibinfo{year}{2018}\natexlab{}.
\newblock \showarticletitle{A review of self-exciting spatio-temporal point
  processes and their applications}.
\newblock \bibinfo{journal}{\emph{Statist. Sci.}} \bibinfo{volume}{33},
  \bibinfo{number}{3} (\bibinfo{year}{2018}), \bibinfo{pages}{299--318}.
\newblock


\bibitem[\protect\citeauthoryear{Shelton, Qin, and Shetty}{Shelton
  et~al\mbox{.}}{2018}]%
        {paper34}
\bibfield{author}{\bibinfo{person}{Christian~R Shelton}, \bibinfo{person}{Zhen
  Qin}, {and} \bibinfo{person}{Chandini Shetty}.}
  \bibinfo{year}{2018}\natexlab{}.
\newblock \showarticletitle{Hawkes process inference with missing data}. In
  \bibinfo{booktitle}{\emph{Thirty-Second AAAI Conference on Artificial
  Intelligence}}.
\newblock


\bibitem[\protect\citeauthoryear{Wang, Fu, Liu, Hu, and Aggarwal}{Wang
  et~al\mbox{.}}{2017a}]%
        {paper9}
\bibfield{author}{\bibinfo{person}{Pengfei Wang}, \bibinfo{person}{Yanjie Fu},
  \bibinfo{person}{Guannan Liu}, \bibinfo{person}{Wenqing Hu}, {and}
  \bibinfo{person}{Charu Aggarwal}.} \bibinfo{year}{2017}\natexlab{a}.
\newblock \showarticletitle{Human mobility synchronization and trip purpose
  detection with mixture of hawkes processes}. In
  \bibinfo{booktitle}{\emph{Proceedings of the 23rd ACM SIGKDD international
  conference on knowledge discovery and data mining}}.
  \bibinfo{pages}{495--503}.
\newblock


\bibitem[\protect\citeauthoryear{Wang, Qin, Pang, Zhang, and Xin}{Wang
  et~al\mbox{.}}{2017b}]%
        {paper17}
\bibfield{author}{\bibinfo{person}{Yan Wang}, \bibinfo{person}{Zongxu Qin},
  \bibinfo{person}{Jun Pang}, \bibinfo{person}{Yang Zhang}, {and}
  \bibinfo{person}{Jin Xin}.} \bibinfo{year}{2017}\natexlab{b}.
\newblock \showarticletitle{Semantic annotation for places in LBSN through
  graph embedding}. In \bibinfo{booktitle}{\emph{Proceedings of the 2017 ACM on
  Conference on Information and Knowledge Management}}.
  \bibinfo{pages}{2343--2346}.
\newblock


\bibitem[\protect\citeauthoryear{Yang, Zhang, Yu, and Yu}{Yang
  et~al\mbox{.}}{2013}]%
        {paper22}
\bibfield{author}{\bibinfo{person}{Dingqi Yang}, \bibinfo{person}{Daqing
  Zhang}, \bibinfo{person}{Zhiyong Yu}, {and} \bibinfo{person}{Zhiwen Yu}.}
  \bibinfo{year}{2013}\natexlab{}.
\newblock \showarticletitle{Fine-grained preference-aware location search
  leveraging crowdsourced digital footprints from LBSNs}. In
  \bibinfo{booktitle}{\emph{Proceedings of the 2013 ACM international joint
  conference on Pervasive and ubiquitous computing}}.
  \bibinfo{pages}{479--488}.
\newblock


\bibitem[\protect\citeauthoryear{Ye, Shou, Lee, Yin, and Janowicz}{Ye
  et~al\mbox{.}}{2011}]%
        {paper3}
\bibfield{author}{\bibinfo{person}{Mao Ye}, \bibinfo{person}{Dong Shou},
  \bibinfo{person}{Wang-Chien Lee}, \bibinfo{person}{Peifeng Yin}, {and}
  \bibinfo{person}{Krzysztof Janowicz}.} \bibinfo{year}{2011}\natexlab{}.
\newblock \showarticletitle{On the semantic annotation of places in
  location-based social networks}. In \bibinfo{booktitle}{\emph{Proceedings of
  the 17th ACM SIGKDD international conference on Knowledge discovery and data
  mining}}. \bibinfo{pages}{520--528}.
\newblock


\bibitem[\protect\citeauthoryear{Yuan, Li, Bertozzi, Brantingham, and
  Porter}{Yuan et~al\mbox{.}}{2019}]%
        {paper7}
\bibfield{author}{\bibinfo{person}{Baichuan Yuan}, \bibinfo{person}{Hao Li},
  \bibinfo{person}{Andrea~L Bertozzi}, \bibinfo{person}{P~Jeffrey Brantingham},
  {and} \bibinfo{person}{Mason~A Porter}.} \bibinfo{year}{2019}\natexlab{}.
\newblock \showarticletitle{Multivariate spatiotemporal hawkes processes and
  network reconstruction}.
\newblock \bibinfo{journal}{\emph{SIAM Journal on Mathematics of Data Science}}
  \bibinfo{volume}{1}, \bibinfo{number}{2} (\bibinfo{year}{2019}),
  \bibinfo{pages}{356--382}.
\newblock


\bibitem[\protect\citeauthoryear{Zarezade, Jafarzadeh, and Rabiee}{Zarezade
  et~al\mbox{.}}{2016}]%
        {paper21}
\bibfield{author}{\bibinfo{person}{Ali Zarezade}, \bibinfo{person}{Sina
  Jafarzadeh}, {and} \bibinfo{person}{Hamid~R Rabiee}.}
  \bibinfo{year}{2016}\natexlab{}.
\newblock \showarticletitle{Spatio-Temporal Modeling of Users' Check-ins in
  Location-Based Social Networks}.
\newblock \bibinfo{journal}{\emph{arXiv preprint arXiv:1611.07710}}
  (\bibinfo{year}{2016}).
\newblock


\end{thebibliography}

\end{document}